\documentstyle[12pt]{article}
\def\tag#1{\space}%
\def\text#1{\space}%
\def\stackunder#1#2{\mathrel{\mathop{#2}\limits_{#1}}}%
\begin{document}
\date{Saclay Preprint SPhT 95/078, June 1995}

\author{
Robi Peschanski \\ CEA, Service de Physique Th\'{e}orique,
 \\ F-91191 Gif sur Yvette Cedex, France}
\title{
IS IT POSSIBLE TO UNIFY THE QCD EVOLUTION OF STRUCTURE FUNCTIONS
IN $X$ AND $Q^2$ ?
}
\maketitle

\begin{abstract}
We start from the two existing QCD evolution equations for structure
functions, the BFKL and DGLAP equations, and discuss the theoretical
hints for a unifying picture of the evolution in $x$ and $Q^2.$ The main
difficulty is due to the property of angular ordering of the gluon radiation
driving the evolution and the cancellation of the related collinear
singularities. At the leading $\log\ 1/x$ and leading $\log \ Q^2$
accuracy, we find a unified set of equations satisfying the constraints.
\end{abstract}
%
%
%
%
%
%
%
%
%
%

\section{The two QCD\ evolution equations and the unification problem}

There is a well-known method to obtain the QCD-theoretical predictions for
quark and gluon structure functions measured in Deep-Inelastic Scattering:
the resummation of leading logarithmic contributions at all orders of the
perturbative theory; indeed, the existence of {\it collinear} and {\it %
infrared} singularities in the evaluation of radiation corrections to the
point like lepton-parton scattering leads to effective coupling constants of
order unity, and thus to the need of resummation techniques. Two types of
resummation do exist.

At large{\rm \ }$Q^2=-q^2,${\rm \ }where $q$ is the quadri-momentum
transferred to the target-proton, the {\it collinear} singularities lead to
effective coupling constants of{\rm \ }order{\rm \ }$\alpha _S(Q^2)\ln
\left(Q^2/\Lambda ^2_{QCD}\right).$ The resummation to all orders of
the leading
logarithms $(LLQ^2)$ leads to the well-known ''Altarelli-Parisi'' (or DGLAP)
evolution equations$^{1,2)}.$

At small values of the Bjorken variable $x=Q^2/2p.q$ ($p$ is the
quadri-momentum of the target), a similar problem appears with the soft part
of the gluon radiation, namely an effective coupling constant of order $%
\alpha _S\ln 1/x.$ It is thus necessary to resum the corresponding leading
logarithms $(LL1/x)$ to all orders. This non-straightforward resummation has
been first performed by L.\ Lipatov and collaborators (BFKL) and leads to a
singular behaviour$^{3)}$ of structure functions at small-$x.$ This result
has been recently revived by Hera results on the quark structure function in
the proton at very small-$x$. They have revealed a behaviour in qualitative
agreement with this QCD prediction$^{4)}$.

For both phenomenological and theoretical reasons, it is interesting to
address the problem of unifying the two mentionned equations into a single
scheme. Since Hera experiments$^{5)}$ cover a very large range in $x$ and $%
Q^2,$ it is quite important to have a unified description of the
QCD-evolution of structure functions in the whole $x-$range. Moreover, it
could solve the dependence on initial conditions for the evolution
equations, which has to do with the unknown non-perturbative regime of QCD.

On a more theoretical ground which is of concern in the present paper, it is
to be remarked that the unification problem has already been suggested and
discussed in the past. It has first been noticed that both $LL1/x$ and $%
LLQ^2 $ can be formally taken into account by a suitable combination$^{5)}$
of the evolution kernels. On a more rigorous basis, it has been shown that a
uniform description of the gluon radiation responsible for the evolution of
structure functions in the whole\ $x$-range is possible due to the property
of {\it angular ordering}$^{6)}.$ Within that picture, it can be shown that
the collinear singularities present in all gluon production amplitudes
contribute only in regions satisfying the following kinematical property:

\begin{equation}
Q/x\gg ...\gg \theta _i\gg \theta _{i-1}\gg ...\gg \theta _1  \tag{1}
\end{equation}

\noindent where $\theta _i\approx (q_T)_i/x_i$ are the angles of
the~emitted gluon with respect to the direction of the first emitted gluon
momentum. One can separate two cases:

\noindent i) at fixed $x,$ corresponding to finite (non strongly ordered) $%
x_i,$ one recovers the well-known $q_t$-ordering of the $LLQ^2$ resummation
technique$^{2)}.$

\noindent ii) at small $x$, the gluon momentum fractions $x_i$ are
necessarily strongly ordered $\left( \frac{x_i}{x_{i-1}}\ll 1\right) $ and
thus q$_t$-ordering is not implied by the relations (1) and angular ordering
is expected to contribute to $LL1/x$ singularities, together with the
infrared singularities.

However, it remains an important constraint to be fulfilled by any
unification scheme based on angular ordering. The Lipatov equation is only
recovered provided that these $LL1/x$ singularities related to angular
ordering exactly cancel in the evolution of structure functions $^{6,7)}.$
This cancellation is not valid for other observables such as multiplicities,
average transverse momentum etc... As we shall see now, this stringent
constraint leads to non-trivial consequences on the unified evolution
equations.

Let us first write the Altarelli-Parisi equations in a suitable form for
unification.\ For this sake, we will restrict ourselves to the case of a
fixed coupling constant $\bar {\alpha}_S$ (as for the BFKL derivation) and
consider the double inverse Mellin transform of the singlet $(F_S)$ and
gluon $(F_G)$ structure functions :

\begin{equation}
F_{S,G}(x,Q^2)=\int \frac{d\gamma }{2i\pi }\ e^{\gamma \ln Q^2/\Lambda
^2}\int \frac{dj}{2i\pi }\ e^{(j-1)\ln 1/x}\varphi _{S,G}(j,\gamma ).
\tag{2}
\end{equation}

\noindent The Altarelli-Parisi equations (for fixed $\bar {\alpha }_S$) can
be written in matrix form for $\varphi _S$ and $\varphi _G$ as follows.

\[
\left(
\begin{tabular}{l}
$\varphi _S^{}$ \\
$\varphi _G$%
\end{tabular}
\right) \equiv
\left(
\begin{tabular}{l}
$\varphi _S^{(0)}$ \\
$\varphi _G^{(0)}$%
\end{tabular}
\right)
\ +
\frac{\overline{\alpha }_S}{4\pi \gamma }
\left(
\begin{tabular}{ll}
\begin{tabular}{l}
$
\nu _F^{}$%
\end{tabular}
&
$2n_F\phi _G^F$
\\
$\phi _F^G$
 & $
\ \ \ \nu _G$%
\end{tabular}
\right)
\left(
\begin{tabular}{l}
$\varphi _S$ \\
$\varphi _G$%
\end{tabular}
\right)
,
\]
\noindent where $\left\{ \nu _G,\nu _F,\phi _G^F,\phi _F^G\right\} $ are the
usual $(j$-dependent) Altarelli-Parisi weights$^{2)},$ and $\varphi
_{S,G}^{(0)}$ are the initial conditions.

Now, one has to modify the equation (3) to take into account the BFKL
contribution in the gluon sector. The dominant contribution can be expressed
$^{3,5)}$ as a singularity in the $j$-plane situated at the value
\begin{equation}
j_L=1+\frac{\bar {\alpha }_SN_C}\pi \chi \left( \gamma \right)  \tag{4}
\end{equation}
\noindent where
\[
\chi (\gamma )\equiv 2\psi (1)-\psi (\gamma )-\psi (1-\gamma )\text{ } ; \
\psi (\gamma )\equiv \frac{d\ln \Gamma (\gamma )}{d\gamma }
\]
\noindent is the eigenvalue-function of the BFKL kernel. Indeed, assuming a
simple pole singularity $\varphi _{S,G}\ \propto \ (j-j_L)^{-1}$ and
inserting it in the Mellin transform (2), one gets:
\begin{equation}
F_{S,G}(x,q^2)=\int \frac{d\gamma }{2i\pi }\ e^{\gamma \ln Q^2/\Lambda ^2} \
e^{\bar {\alpha }^S\frac{N_c}\pi \chi (\gamma )\ln 1/x}\simeq \left( _{%
\overline{\Lambda }^2}^{Q^2}\right) ^{1/2}x^{-\overline{\alpha }_S\frac{N_c}%
\pi4 \ln 2},  \tag{5}
\end{equation}
\noindent where one has made use of a saddle point method to integrate
around the point $\gamma _c=1/2,\ \chi (\gamma _c)=4\ln 2$. Equation (5)
corresponds exactly to the leading BFKL behaviour$^{3)},$ up to logarithmic
corrections (which would be determined by the nature of the singularity in
the $j$-plane).

A consistent modification of equation (3) in order to complement the
singular behaviour (4) is the following$^{8)}$; let us replace the gluon
contribution to the anomalous dimensions:
\begin{equation}
\nu _{G(j)}\longrightarrow \nu _{G(j)}^{*}=\gamma\ \chi (\gamma )\left\{ \nu
_G+\Psi \right\} -\Psi ,  \tag{6}
\end{equation}
\noindent where $\Psi $ is an arbitrary function holomorphic in the $j$%
-plane near $j=1$ and below. Such a modification inserted in equation (3)
provides a formal unification of the Altarelli-Parisi and Lipatov kernels.
Indeed, inverting the relation (3), after the replacement $\nu
_G\longrightarrow \nu _G^{*}$, one gets

\[
\left(
\begin{tabular}{l}
$\varphi _G^{}$ \\
$\varphi _S$%
\end{tabular}
\right) \equiv \frac 1{D(j,\gamma )}\left(
\begin{tabular}{ll}
\begin{tabular}{l}
$1-\frac{\overline{\alpha }_S}{4\pi \gamma }\ \nu _F^{}$%
\end{tabular}
& $\ \ \ \ \frac{\overline{\alpha }_S}{4\pi \gamma} \ \phi _F^G$ \\
$\frac{\overline{\alpha }_S}{4\pi \gamma }2n_F\ \phi _G^F$ & $1-\frac{%
\overline{\alpha }_S^{}}{4\pi \gamma }\ \nu _G^{*}$%
\end{tabular}
\right) \left(
\begin{tabular}{l}
$\varphi _G^{(0)}$ \\
$\varphi _S^{(0)}$%
\end{tabular}
\right) ,
\]
\noindent with
\begin{equation}
D(j,\gamma )=1-\frac{\overline{\alpha }}{4\pi \gamma }\left(
\nu ^{*}_G+\ \nu _F\right) +\left( \frac{\overline{\alpha }}{4\pi \gamma }%
\right) ^2\left( \nu _F\nu _G^{*}-2n_F\phi _G^F\phi _F^G\right) .  \tag{7}
\end{equation}
\noindent Now, the solutions of equation~(6) depend on the region in the
complex $j$-plane involved in the Mellin transform (2), and thus on the
region in $\ln 1/x$ one is investigating; two cases appear:

\noindent i) when $x$ is not small, $\overline{\alpha }_S\ln 1/x\ll 1,$ the
modification (5) has no effect, since the zeroes of $D(j,\gamma )$ are
obtained for small values of $\gamma $ (of order $\overline{\alpha }_S).$ In
that limit, one has from the very definition of $\chi (\gamma ): $

\begin{equation}
\chi (\gamma )\approx 1/\gamma +{\cal O}(\gamma ^2);\ \nu _G^{*}\approx \nu
_G+{\cal O}(\overline{\alpha }^3).  \tag{8}
\end{equation}

\noindent One recovers the ordinary Altarelli-Parisi scheme$^{3)}$ and the
corresponding evolution equations (at fixed $\overline{\alpha }_S).$

\noindent ii) When $\overline{\alpha }_S\ln 1/x={\cal O}(1),$ the singular
structure of the BFKL kernel plays a role, driving the relevant domain of
the Mellin integration over $\gamma $ near to the ''critical'' value $\gamma
_c=1/2.$

\noindent In those conditions one recovers the singular behaviour compatible
with the BFKL calculations. Taking the appropriate limit $j\rightarrow 1,$ $%
\overline{\alpha }_S/(j-1)={\cal O}(1):$

\begin{equation}
D(j,\gamma )\ \alpha \ 1-\frac{\overline{\alpha }N_C \chi (\gamma )}{4\pi
(j-1)}\stackunder{j\longrightarrow 1}{\approx }1-\frac{\overline{\alpha }}%
\pi N_C\frac{4\ln 2}{j-1}  \tag{9}
\end{equation}

At first sight, the DGLAP and BFKL evolution equations can be unified for an
arbitrary regular function $\Psi (j)$ in eq.(6). For instance let us
consider the combination:
\begin{equation}
\nu _G^{*}+\nu _F=\gamma \chi (\gamma )\left\{ \nu _G+\Psi \right\} +\nu
_F-\Psi  \tag{8}
\end{equation}

\noindent appearing in $D(j,\gamma )$ at first order in $\overline{\alpha }%
_S $. Following the arguments of refs.$^{6,7)}$ as we have stressed upon in
our introductory discussion, the quark-loop contribution $\nu _F,$ which is
present at fixed value of $x$ as a result of collinear singularities, should
be absent from the evolution equations for small value of $x.$ More
precisely, if not cancelled appropriately, it would bring a new $LL1/x$
singularity, due to the angular-ordering property including emitted quarks.
It is thus compelling to choose $\nu _F
\approx \Psi $  when
$
j\rightarrow 1
$
in order to obtain the desired cancellation. This is just
the mechanism proposed in our paper$^{8)}.$ Indeed, the problem has been
noticed to arise when one is to include ''finite parts'' into the evolution
equations at small-$x^{9)}.$

As a consequence, considering the proposed cancellation to be valid in the $%
j $-plane around the leading singularity $j_L,$ one writes
\begin{eqnarray}
\Psi (j_L) &\approx &\nu _F(j_L)  \tag{10} \\
\nu _G(j_L)+\nu _F(j_L) &=&\left[ \frac{\overline{\alpha }N_C \log 2}\pi
\right] ^{-1}.  \nonumber
\end{eqnarray}

\noindent As noticed in Ref.[8], the equations (10) lead to an appreciable
modification of the location $j_L$ of the BFKL singularity endpoint, in
better agreement with phenomenological determinations$^{4)}.$

\noindent Among other interesting properties, the set of equations (10)
ensures (at first order in $\overline{\alpha }_S$) that the quark loops do
not contribute to the small-$x$ evolution. Moreover, the expression of $%
D(j,\gamma $ ) preserves the position of the saddle point in the $%
\gamma $ plane at the critical value $1/2,$ as expected from the conformal
properties of the BFKL kernel$^{10)}.$

As a conclusion, the unification of the evolution equations for structure
functions appears possible, at least in the leading logarithmic
approximation and at fixed coupling constant $\overline{\alpha }_S.$ Despite
stringent constraints due to the mismatch of $LL1/x$ and $LLQ^2$
perturbative resummations, a unified set of equations for the whole range in $%
x$ and $Q^2$ can be written and leads to non-trivial predictions. A number
of interesting questions remain open for future investigation, let us list
some of them:\smallskip\

\noindent 1) Is it possible to implement unified evolution equations at the
next-leading-order ?\smallskip\

\noindent 2) In the same context, how the result might be  influenced by the
running of $\alpha _S$ ?\smallskip\

\noindent 3) What are the phenomenological consequences of unified equations
?\smallskip\

\noindent We hope to be able to provide answers to these questions in the
near future.\smallskip\

\section{Acknowledgments}

Samuel Wallon has been associated with the most part of what is written in
the present note.\hspace{1.0in}

\section{References}

\noindent  1) G.\ Altarelli and G.\ Parisi,{\it \ Nucl.\ Phys.}\underline{%
{\it \ B126}} (1977) 293; V.N. Gribov and L.N. Lipatov,{\it \ Sov. Journ.
Nucl. Phys.}\ (1972) 438 and 675; Yu.L. Dokshitzer,{\it \ Sov.\ Phys.\ JETP
\underline{46} }(1977) 641.\smallskip\

\noindent  2) For a review on perturbative QCD: (not including BFKL) and
notations used in the paper: {\it Basics of perturbative QCD,} Yu.\ L.\
Dokshitzer, V.A. Khoze, A.N. Mueller and S.I. Troyan (J. Tran Than Van ed.
Editions Fronti\` eres) 1991.\smallskip\

\noindent  3) E.A. Kuraev, L.N. Lipatov, V.S. Fadin, {\it Sov.\ Phys. JETP }%
\underline{45} (1977) 199; Ya.Ya. Balistsky and L.N. Lipatov,{\it \ Sov.
Nucl. Phys. }\underline{28} (1978) 822.\smallskip\

\noindent  4) H1 Coll. {\it Nucl.\ Phys. }\underline{B 407} (1993) 515,
\underline{B 439} (1995) 471. \ Zeus Coll., {\it Phys.\ Lett. }\underline{B
316} (1993) 412; DESY preprint 94-143, to be published in {\it Zeit.} {\it %
F\"{u}r Phys.} (1995).\smallskip\

\noindent  5) L.V. Gribov, E.M. Levin and M.G. Ryskin, {\it Zh. Eksp.,
Teor., Fiz} \underline{80} (1981) 2132; {\it Phys.\ Rep. }\underline{100}
(1983) 1.\smallskip\

\noindent  6) M. Ciafaloni, {\it Nucl. Phys.} \underline{B 296} (1987) 249;
\ S.\ Catani, F. Fiorani and G. Marchesini, {\it Phys.\ Lett. }\underline{B
234} (1990) 389 and {\it Nucl. Phys.} \underline{B 336} (1990) 12; S.Catani,
F.\ Fiorani, G.\ Marchesini and G. Oriani, {\it Nucl.\ Phys.\ \underline{B
361} (1991) 645.}\smallskip\

\noindent  7) C.\ Marchesini, ''QCD coherence in the structure function and
associated distributions at small $x$'', Milano preprint IFUM 486-FT (1994).%
\smallskip\

\noindent  8)\ \thinspace R. Peschanski and S.\ Wallon, {\it Phys. Lett.}
\underline{B 349} (1995) 357.\smallskip\

\noindent  9) G.\ Marchesini, contribution to the Workshop on Deep Inelastic
Scattering and QCD, DIS 95', Paris, April 19-24, 1995.\smallskip\

\noindent  10) L. Lipatov, {\it Phys.\ Lett. }\underline{B 309} (1993) 393,
and references therein.

\end{document}